\documentclass[twocolumn,prl,amssymb,floatfix]{revtex4}
\usepackage[latin9]{inputenc}
\usepackage{amsmath}
\usepackage{amssymb}
\usepackage{graphicx}

\makeatletter
\@ifundefined{textcolor}{}
{%
 \definecolor{BLACK}{gray}{0}
 \definecolor{WHITE}{gray}{1}
 \definecolor{RED}{rgb}{1,0,0}
 \definecolor{GREEN}{rgb}{0,1,0}
 \definecolor{BLUE}{rgb}{0,0,1}
 \definecolor{CYAN}{cmyk}{1,0,0,0}
 \definecolor{MAGENTA}{cmyk}{0,1,0,0}
 \definecolor{YELLOW}{cmyk}{0,0,1,0}
}

\usepackage{amsfonts}\usepackage{bbm}\usepackage{epsfig}\usepackage{graphics}

\newcommand{\gsim}{\,\lower2truept\hbox{${>\atop\hbox{\raise4truept\hbox{$\sim$}}}$}\,}

\newcommand{\be}{\begin{equation}}
\newcommand{\ee}{\end{equation}}
\newcommand{\bea}{\begin{eqnarray}}
\newcommand{\eea}{\end{eqnarray}}

\makeatother

\begin{document}

\title{Friction in Gravitational waves: a test for early-time modified gravity}

\author{Valeria Pettorino and Luca Amendola}

\affiliation{Institut für Theoretische Physik, Universität Heidelberg, Philosophenweg
16, D-69120 Heidelberg}
\begin{abstract}
Modified gravity theories predict in general a non standard equation
for the propagation of gravitational waves. Here we discuss the impact
of modified friction and speed of tensor modes on cosmic microwave
polarization B modes. We show that the non standard friction term,
parametrized by $\alpha_{M}$, is degenerate with the tensor-to-scalar
ratio $r$, so that small values of $r$ can be compensated by negative
constant values of $\alpha_{M}$. We quantify this degeneracy and
its dependence on the epoch at which $\alpha_{M}$ is different from
the standard, zero, value and on the speed of gravitational waves
$c_{T}$. In the particular case of scalar-tensor theories, $\alpha_{M}$
is constant and strongly constrained by background and scalar perturbations,
$0\le\alpha_{M}<0.01$ and the degeneracy with $r$ is removed. In
more general cases however such tight bounds are weakened and the
B modes can provide useful constraints on early-time modified gravity. 
\end{abstract}
\maketitle
In Modified Gravity models, the equation for the tensor metric perturbations
(gravitational waves) is affected in several ways. First, the speed
$c_{T}$ of the gravitational waves can be different from the speed
of light\cite{amendola_ballesteros_pettorino_2014,2014arXiv1405.7974R}.
A second effect is instead related to a modification of the friction
term in the tensor equation that depends on the evolution rate of
the effective Planck mass or equivalently on the effective universal
gravitational interaction of the cosmological model \cite{Bellini:2014fua}.
In more general modifications of gravity, for instance in bimetric
models \cite{Hassan:2011vm}, two coupled tensor equations are present,
corresponding to the two metrics of the theory and additional changes
are possible \cite{Saltas:2014dha}. Considering only the first two
possible modifications, gravitational wave speed and friction, the
linear generalized tensor equation for the amplitude $h$ in vacuum
can be written in a Friedmann-Robertson-Walker (FRW) metric as \cite{DeFelice:2011bh,Bellini:2014fua}:
\begin{equation}
\ddot{h}+(3+\alpha_{M})H\dot{h}+c_{T}^{2}\frac{k^{2}}{a^{2}}h=0\,\,\,,\,\label{eq:gw-1}
\end{equation}
where the dot represents derivative with respect to cosmic time, $\alpha_{M},c_{T}$
are time-dependent functions that vary with the specific model, $k$
is the wavenumber, $a$ is the scale factor and $H$ the Hubble function.
In the standard case one has $\alpha_{M}=0$ and the speed of gravitational
waves $c_{T}$ equals the speed of light, $c_{T}=1$. General models
belonging to the so-called Horndeski Lagrangian \cite{Horndeski:1974wa}
produce both effects, i.e. $\alpha_{M}\not=0$ and $c_{T}\not=1$.
When anisotropic stress is present \cite{2013PhRvD..88h4008D}, a source term in (\ref{eq:gw-1}) is also
included, although it is typically negligible.

Any modification of the tensor wave equation can potentially lead
to observable effects on the Cosmic Microwave Background (CMB), on
both the temperature and the polarization spectra. The recent measurement
of the B-modes at multipoles around $\ell=100$ reported by the BICEP2
experiment \cite{Ade:2014xna}, as well as follow up analysis on foreground
contributions such as dust emission \cite{2012arXiv1212.5225B,Liu:2014mpa,2014arXiv1405.0874P, 2014arXiv1409.5738P},
has motivated a great interest in the information contained in the
polarization B-mode signal, especially for as concerns the inflationary
dynamics. As it is well known, in absence of vector sources, the primordial
B-mode spectrum is generated exclusively by tensor waves and affects
small to intermediate multipoles. Larger multipoles $\ell\gsim100$
are mainly affected by CMB-lensing. In \cite{amendola_ballesteros_pettorino_2014,2014arXiv1405.7974R}
it has been shown that the gravitational wave speed at the epoch of
decoupling or before affects the position of the inflationary and
of the reionization peak in the polarization B-modes so that a measurement
of B-modes at $\ell\approx100$ can be employed to set limits on the
early-time speed of gravitational waves.

In many cases, the effects of these changes on the CMB can be safely
neglected if one assumes that gravity deviates from the Einsteinian
form only recently, as in several models proposed to explain the recent
epoch of cosmic acceleration by non-standard gravity. In general,
however, modification of gravity can occur at any time in the past;
in some models, e.g. Brans - Dicke theories or some bimetric models
\cite{Konnig:2014xva}, gravity is modified at all times. In this
Letter we wish to study the impact and limits that the current observations
of B-modes can set on the two modified gravity tensor parameters,
$\alpha_{M}$ and $c_{T}$. Although they are both in general time-dependent
quantities, we assume here for simplicity that they are constant or
that they deviate from the standard case only at early time, i.e.
before some epoch $z_{d}$.

\begin{figure}[ht]
\centering \includegraphics[width=0.45\textwidth]{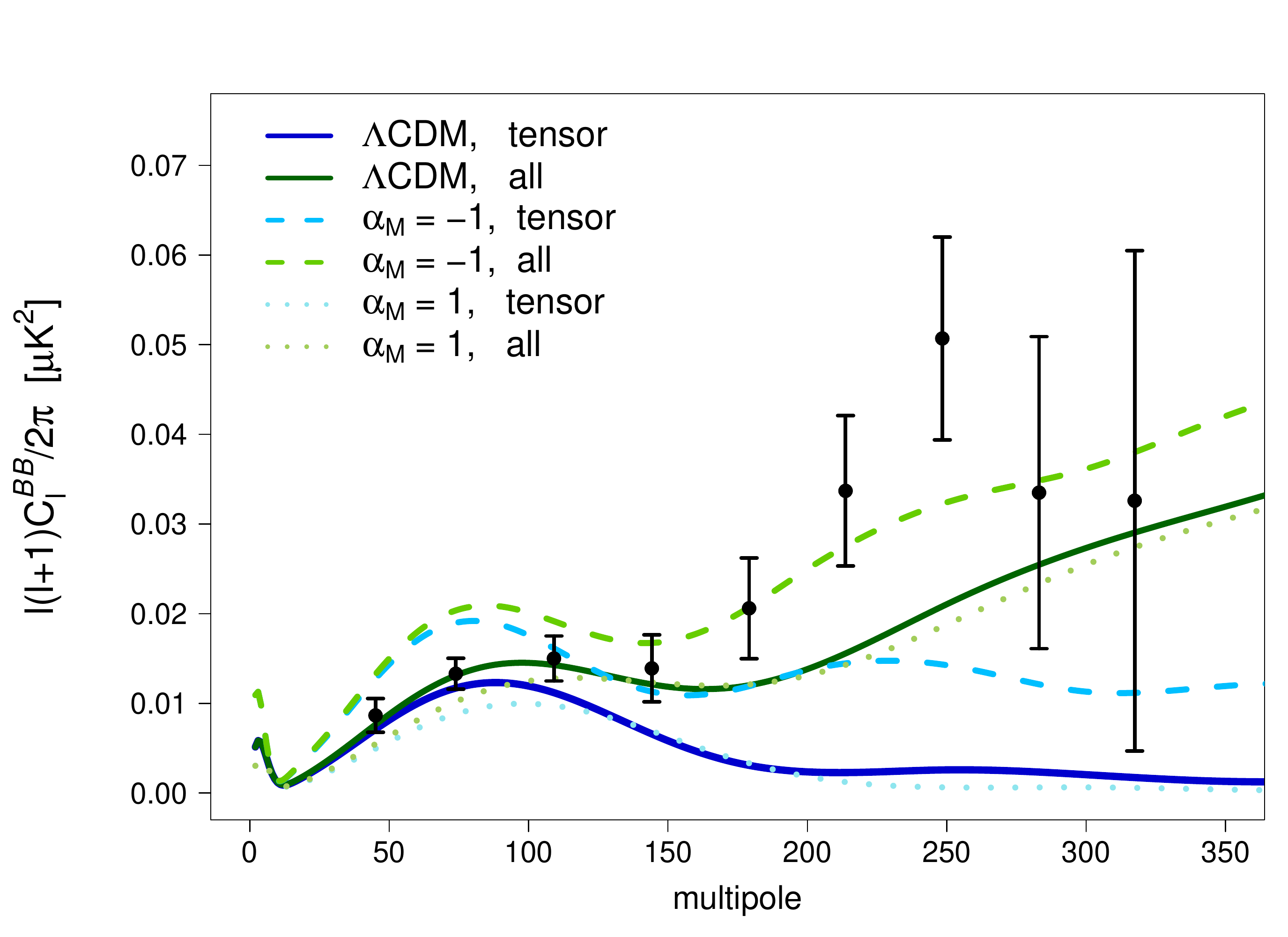}
\protect\caption{CMB BB spectra for two values of $\alpha_{M}$ and for $\Lambda$CDM.
We plot both tensor contribution only (blue lines) and the total spectra
(green lines), including lensing. The data points are from BICEP2.
The case $\alpha_{M}=0$ coincides with $\Lambda$CDM. For this plot,
$c_{T}^{2}=1$ and $r_{0.05}=0.2$.}
\label{fig_1} 
\end{figure}

We have modified the tensor equation in CAMB %
\footnote{http://camb.info/%
} and combined it within CosmoMC \cite{cosmomc_lewis_bridle_2002}
to include the $\alpha_{M},c_{T}$ parameters. We first consider the
case in which $c_{T}=1$ but $\alpha_{M}$ is arbitrary and constant.
All the other parameters are as in standard $\Lambda$CDM. In Fig.(\ref{fig_1})
we show the effect of $\alpha_{M}$ on the BB spectrum of the CMB
both on the tensor modes only and on the total spectrum. As expected,
a positive $\alpha_{M}$ increases the friction term and therefore
reduces the wave amplitude, while a negative $\alpha_{M}$ has the
opposite effect.

\begin{figure}[ht]
\centering \includegraphics[width=0.45\textwidth]{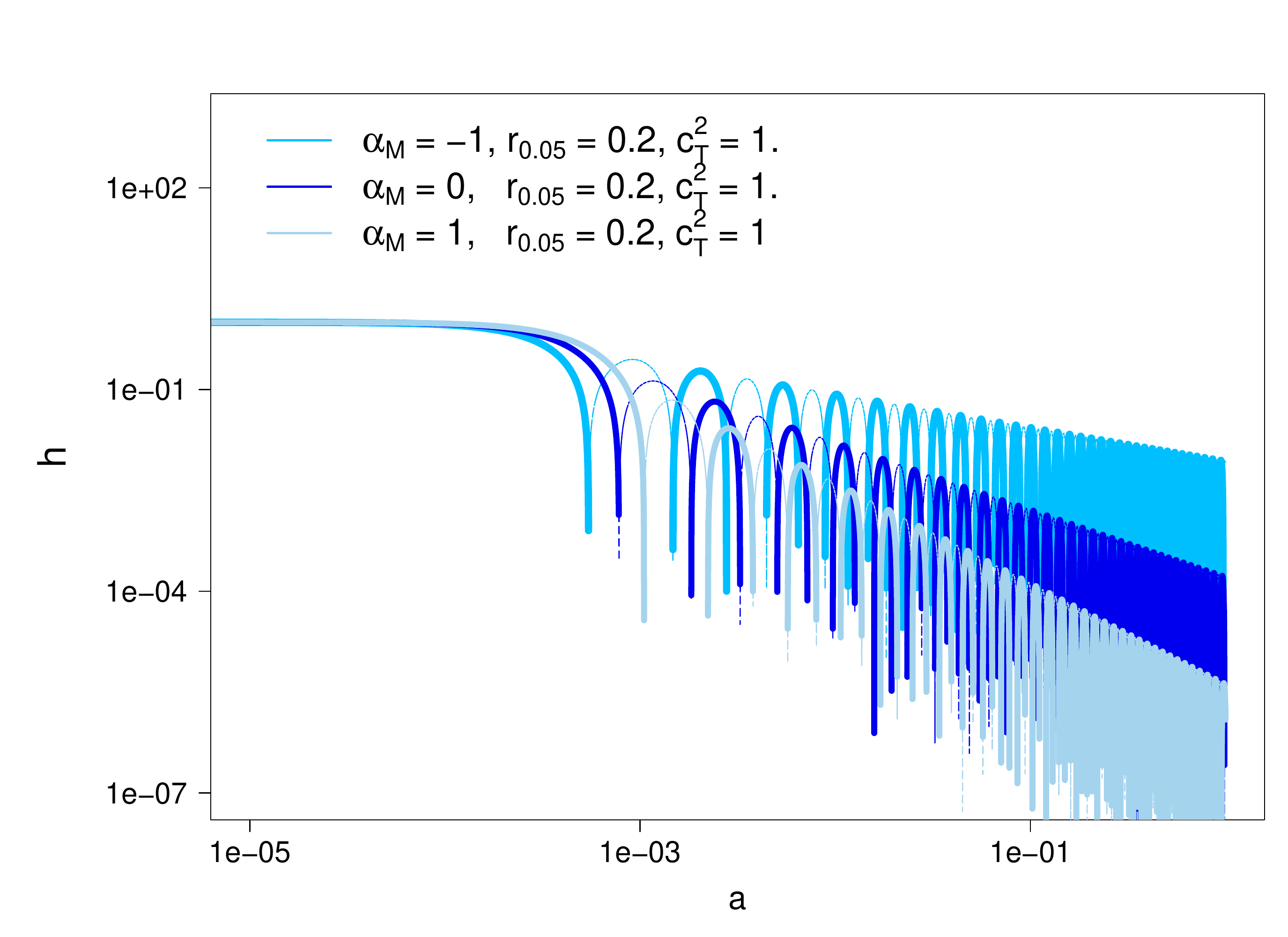}
\protect\caption{Tensor perturbation evolution for $\rm{k} \sim 0.01$ for two values of $\alpha_{M}$ and for $\Lambda$CDM.
Thin dashed lines show the absolute value.}
\label{hgravity_plot} 
\end{figure}

\begin{figure}[ht]
\centering \includegraphics[width=0.4\textwidth]{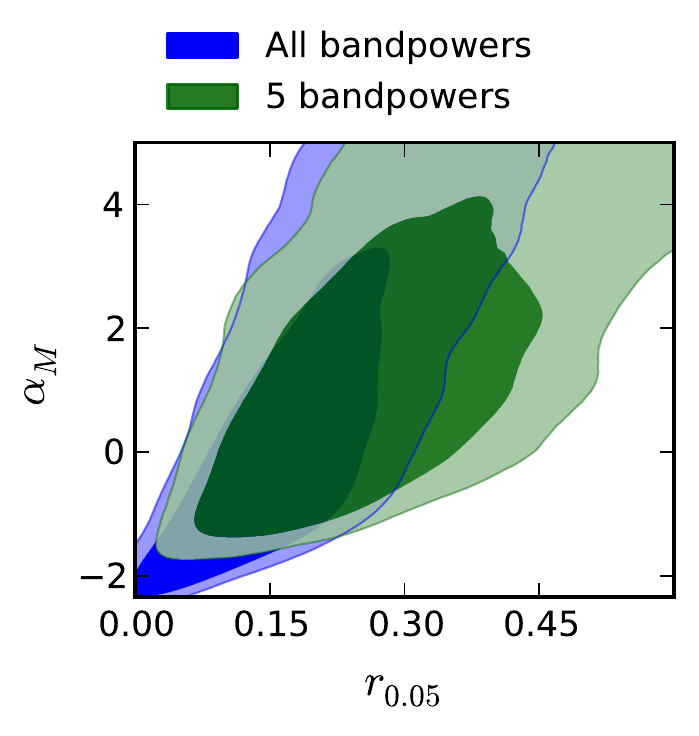}
\protect\caption{Posterior likelihood for $\alpha_{M}$ and $r_{0.05}$. Blue contours
are obtained using all nine bandpowers from BICEP2 while green (top)
contours include only the first five bandpowers.}

\label{contours-am} 
\end{figure}

\label{sec:speed} From Fig.(\ref{fig_1}) we can expect a degeneracy
between the tensor-to-scalar ratio $r$ and $\alpha_{M}$, as they
both regulate the amplitude of the primordial peak. Comparing only
with the BICEP2 data and fixing the optical depth to the Planck best
fit value ($\tau=0.09$ for Planck + WMAP polarization \cite{Planck_params}),
we obtain the allowed region for $(\alpha_{M},r)$, shown in Fig.(\ref{contours-am}),
which clearly shows the degeneracy. Values of $r$ close to zero can
be reconciled with the \emph{ }BICEP2 data if $\alpha_{M}$ is close
to $-2$. This is the central result of this paper. In the same figure
we also compare the results obtained when including all nine band
powers of BICEP2 with the case in which only the first five are included.
As expected from Fig.(\ref{fig_1}), a negative value of $\alpha_{M}$
also increases tensor modes at large multipoles, therefore smaller
values of $\alpha_{M}$ are favoured if also the last four (higher
multipole) band powers of BICEP2 are included.
The corresponding evolution of tensor perturbations is shown in Fig.(\ref{hgravity_plot}).

Before drawing any conclusion from Fig.(\ref{contours-am}), one should
consider however that the term $\alpha_{M}$, as already mentioned,
is proportional to the time derivative of $G_{{\rm {eff}}}$. It enters,
therefore, into the scalar perturbation equations and contributes
to the Sachs-Wolfe, Integrated Sachs-Wolfe and lensing signal, affecting
both the temperature and the polarization spectra. In principle, one
should therefore consider all the spectra at the same time, and also
the background evolution which in general will be different from $\Lambda$CDM.
However to a large extent the low-$\ell$ B modes are independent
of all the other signals (T and E spectra, and high-$\ell$ B modes)
since they depend uniquely on the tensor modes; on the other hand,
tensor modes affect only marginally the other CMB spectra. Therefore,
to a first approximation, we can just use the already available constraints
on $\alpha_{M}$ in some specific model to see which fraction of the
parameters space of Fig.(\ref{contours-am}) is allowed.

\begin{figure}[ht]
\centering \includegraphics[width=0.4\textwidth]{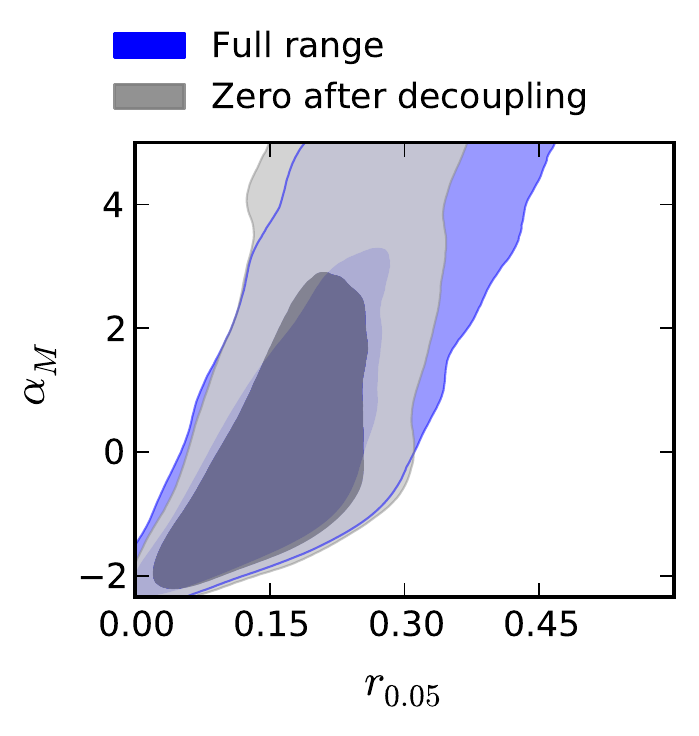}
\protect\caption{Posterior likelihood for $\alpha_{M}$ and $r_{0.05}$. Blue (top)
contours are obtained using all nine bandpowers from BICEP2 and modifying
$\alpha_{M}$ in the full $z$ range (same as previous figures). Gray
contours correspond to the case in which $\alpha_{M}$ is only modified
up to decoupling. We fix $c_{T}=1$.}

\label{contours-am_decoupling} 
\end{figure}

Let us consider for instance one of the simplest cases of modified
gravity, the scalar-tensor theory. Perturbation equations for scalar-tensor
theories with a scalar field $\phi$ have been calculated for example
in \cite{Amendola:1989vj,Hwang:1990re} including tensor equations
for the metric. For a model with Lagrangian 
\begin{equation}
L=\sqrt{-g}[f(\psi)R-2\kappa^{2}(\frac{1}{2}\psi_{;\mu}\psi^{;\mu}+V(\psi))]
\end{equation}
where $\kappa^{2}=8\pi G$ and $R$ is Ricci's scalar, the gravitational
wave equation turns out to be 
\begin{equation}
\ddot{h}+\left(3H+\frac{\dot{\phi}}{\phi}\right)\dot{h}+\frac{k^{2}}{a^{2}}h=0
\end{equation}
if we introduce the rescaled field $\phi=f(\psi)$. Therefore in scalar-tensor
theories we can readily identify 
\begin{equation}
\alpha_{M}=\frac{\dot{\phi}}{\phi H}=\frac{d\log\phi}{d\log a}
\end{equation}
In the simplest form of scalar-tensor model, the original Brans-Dicke
model
\begin{equation}
L=\sqrt{-g}[\phi R-\kappa^{2}\frac{\omega}{\phi}\phi_{;\mu}\phi^{;\mu}]
\end{equation}
the evolution is controlled by the single observable parameter $\omega$.
As long as the matter density is dominated by a component with constant
equation of state $w_{m}$ the background expansion has a simple analytical
solution \cite{1969PThPh..42..544N}, 
\begin{equation}
\phi\sim t^{\frac{2-6w_{m}}{4+3\omega-3\omega w_{m}^{2}}},\quad a\sim t^{\frac{2+2\omega-2\omega w_{m}}{4+3\omega-3\omega w_{m}^{2}}}
\end{equation}
In this case, we can relate the $\alpha_{M}$ parameter to $\omega$,
so that: 
\begin{equation}
\alpha_{M}=\frac{2-6w_{m}}{2+2\omega-2\omega w_{m}}\,\,\,\,.
\end{equation}
During the matter dominated era, this becomes simply $\alpha_{M}=1/(1+\omega)$.
The local gravity constraints on $\omega$ are extremely tight, $\omega>25,000$
\cite{Agashe:2014kda}, but they could be escaped if the scalar couples
to dark matter only. In order to obtain more general constraints one
can use cosmological observations, as e.g. CMB or large-scale structure.
In this case one has typically $\omega>100$ (see e.g. \cite{2005PhRvD..71j4025A}),
so we get $0\,\le\,\,\alpha_{M}<0.01$. Taking this constraint at
face value, we should conclude that the effect of $\alpha_{M}$ on
the tensor modes is practically negligible. However, this is only
true for the particular case in which $\alpha_{M}={\rm {const}}$
at all times. If $\alpha_{M}$ varies in time (as expected in general)
and in particular if $\alpha_{M}$ is very small after decoupling,
the scalar effects can become arbitrarily weak since they are mostly
due to post-decoupling physics (except of course for the inflationary
initial conditions, that we are assuming to be independent of the
gravity modifications we are considering here). On the other hand,
B modes depend on the evolution of gravitational waves before decoupling.
To give an example, if we have an extreme case in which $\alpha_{M}$
suddenly decreases to zero just after decoupling then the B modes
would be practically the same as if $\alpha_{M}$ were constant at
all times (see Fig. \ref{contours-am_decoupling}), while the scalar
perturbations would be the same as in $\Lambda$CDM. In this case,
$\alpha_{M}$ has effect on B modes up to decoupling, while it has
no impact on secondary anisotropies such as integrated Sachs-Wolfe
and CMB lensing. As a consequence, the constraints we obtain on $\alpha_{M}$
from B modes are also valid for all those models in which the modified
gravity effects are due to an $\alpha_{M}$ that is a non zero constant
only until the epoch of decoupling. Needless to say, assuming $\alpha_{M}$
to be exactly zero immediately after the decoupling serves merely
as an illustrative example and should not be taken as a realistic
model.

A larger friction term has two competing effects: on one side, it
delays the horizon reenter and the subsequent damping, therefore momentarily
enhancing the tensor amplitude; on the other, it increases the damping
itself, quenching the ``acoustic'' oscillations more than in the
standard case. This implies that if the epoch at which $\alpha_{M}$
goes to zero moves from the decoupling to an earlier epoch, eg. $z_{d}=2000$,
the BB spectra change in a non trivial way. High multipoles, e.g.
$\ell\gg100$, which correspond to wavelengths that are well within
the horizon at decoupling, move monotonically closer to the $\Lambda$CDM
spectrum the higher is $z_{d}$, as expected since $\alpha_{M}$ vanishes
during a longer part of the evolution. Modes that are crossing the
horizon at decoupling, however, have a more complicate behavior since
they are just beginning their oscillations: for these scales the trend
is not monotonic with $z_{d}$ when $z_{d}$ is close to the decoupling
epoch. Only for $z_{d}$ larger than 10,000 are they back to the $\Lambda$CDM
amplitude. Since $r$ is related to the primordial peak in the BB
spectrum, which changes non monotonically if $\alpha_{M}$ is set
to zero before decoupling, this also means that the direction of the
degeneracy between $r$ and $\alpha_{M}$ is not necessarily the one
in Fig. (\ref{contours-am}) if $\alpha_{M}$ is not constant at all
times until decoupling. {In particular, even for negative $\alpha_{M}=-1$,
the primordial B spectrum becomes smaller than $\Lambda$CDM if $z_{d}\approx1500$.
The degeneracy becomes then inverted (larger $r$ corresponding to
smaller $\alpha_{M}$ if $z_{d}$ is larger than 1500, for instance
fixed to matter-radiation equality).} The degeneracy between $r_{{0.05}}$ and $n_s$ shown in \cite{amendola_ballesteros_pettorino_2014} is weakened when $\alpha_M$ is free to vary.

Since $\alpha_{M}$ is related to $G_{{\rm {eff}}}$, another caveat
is that $G_{{\rm {eff}}}$ influences the growth of scalar perturbations
before decoupling. As a consequence, $r$ can also change indirectly
through the Poisson equation and the modification of scalar perturbations,
even if tensor spectra stay the same. This effect will have to be
investigated more systematically implementing also the full set of
scalar equations, which is beyond the scope of this paper. However
we note that it will mainly play a role only after perturbations have
entered the horizon, affecting scalar perturbations only for $\ell\gsim100$
through a change in the Sachs Wolfe effect, given by the different
gravitational potential. Moreover, a change in $G_{{\rm {eff}}}$
could be compensated with a change in $\Omega_{m}$. If $z_{d}$ is
set to a redshift before the equality, the effect should be negligible
as matter perturbations only start growing when $\alpha_{M}$ is already
zero.

\begin{figure}[ht]
\centering \includegraphics[width=0.4\textwidth]{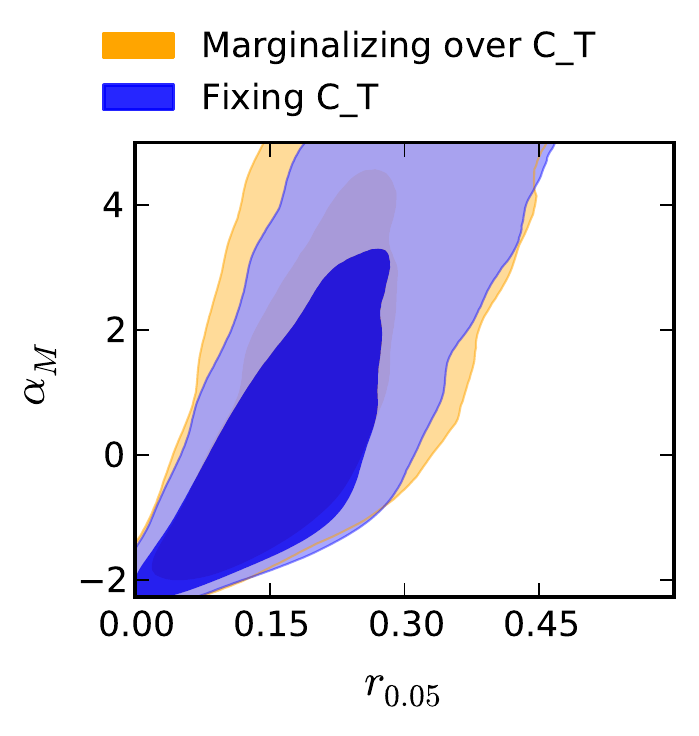}
\protect\caption{Posterior likelihood for $\alpha_{M}$ and $r_{0.05}$. Blue (top)
contours are obtained using all nine bandpowers from BICEP2 and fixing
$c_{T}$ while orange contours marginalize over $c_{T}$.}

\label{contours-am_ct} 
\end{figure}

\begin{figure}[ht]
\centering \includegraphics[width=0.35\textwidth]{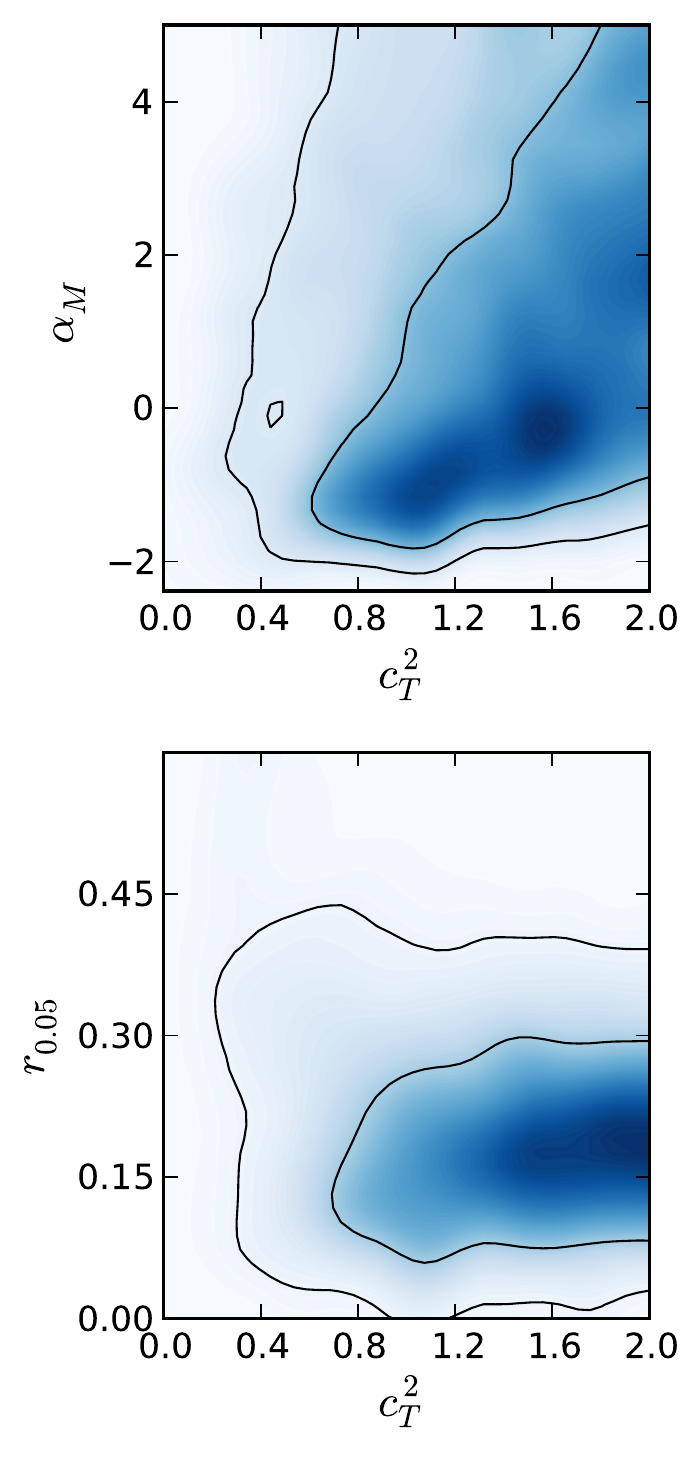}
\protect\caption{Posterior likelihood for $\alpha_{M}$ and $r_{0.05}$ vs $c_{T}^{2}$.}

\label{contours-am_ct_degeneracy} 
\end{figure}

We can now consider simultaneously both parameters, $\alpha_{M}$
and $c_{T}$. In this case, marginalizing over $c_{T}$, we obtain
the contours in Fig. (\ref{contours-am_ct}), which appear to be very
similar to the case in which we fix $c_{T}=1$. The contours of $\alpha_{M},c_{T}$
and $r,c_{T}$ are in Fig. (\ref{contours-am_ct_degeneracy}). We
find a mean and standard deviation for $c_{T}^{2}=1.3\pm0.5$ with
a best-fit of $c_{T}^{2}({\rm {best\, fit})=0.8}$ which is compatible
with $\Lambda$CDM and, most of all, the same value found in \cite{amendola_ballesteros_pettorino_2014,2014arXiv1405.7974R}
using also the temperature power spectra besides BICEP2. This confirms
that the relevant epoch during which the tensor equation is affected
mainly by B modes is the one before decoupling.

We remark that the speed of gravitational waves can be constrained
also with the gravi-Cherenkov effect (see e.g. \cite{1980AnPhy.125...35C,2012JCAP...07..050K,Moore:2001bv}),
which gives a tight lower limit but no upper limit. However, this
methods applies only locally (or at most within the distance scale
of cosmic rays) and/or at the present time; therefore, they are complementary
to the observation of B-modes. 

Finally, we note that the theoretical BB spectrum shows another peak
at $\ell\approx5$, still to be detected, due to the effects of tensor
modes on the scattering during reionization. A non-zero $\alpha_{M}$
changes the amplitude of the reionization peak, similarly to what
happens when $c_{T}$ is modified \cite{amendola_ballesteros_pettorino_2014}.
Its detection, for instance with the proposed satellite mission LiteBIRD
\cite{2013arXiv1311.2847M} %
\footnote{http://litebird.jp/%
}, could therefore put constraints on the friction and gravitational
wave speed before and during reionization.

In conclusion, we have studied the impact on B modes of a modified
gravity tensor equation taking into account both the friction and
the speed term. We have shown that a low value of the tensor-to-scalar
ratio $r$ can be reconciled with the BICEP2 recent data if $ $$\alpha_{M}$
is close to -2. In the case in which all signal will turn out to be mainly due to dust \cite{2014arXiv1409.5738P}, 
the BICEP2 measurements would mainly give a bound on the B mode primordial signal, that can still be used to constrain modifications of gravity. The lensing part of the B mode signal is also degenerate with modifications of gravity, as shown in Fig.(\ref{fig_1}). In specific models, such as Brans Dicke scalar-tensor
theories, the $\alpha_{M}$ parameter is already strongly constrained
by the temperature spectra and the degeneracy with $r$ is removed.
We argue however that in general this is not the case and therefore
the BB spectrum is a useful test of early time modifications of gravity.
\\
 
\begin{acknowledgments}
The authors acknowledge the DFG TransRegio TRR33 grant on `The Dark
Universe'. The authors thank Guillermo Ballesteros, Francesco Montanari
and Marco Raveri for fruitful discussion. \\
 \\

\end{acknowledgments}
\bibliographystyle{plain}
\bibliography{joint}

\end{document}